\documentclass{moriond}

\def\be{\begin{equation}}
\def\ee{\end{equation}}
\def\bea{\begin{eqnarray}}
\def\eea{\end{eqnarray}}

\newcommand{\Photo}{}

\newcommand{\gev}{~\mathrm{GeV}}

\def\reffi#1{\mbox{Fig.~\ref{#1}}}

\begin{document}
\begin{flushright}
KA-TP-05-2023
\end{flushright}
\vspace*{4cm}
\title{Scalar extensions of the SM and recent experimental anomalies}

\author{T.~Biek\"otter}

\address{Institute for Theoretical Physics, Karlsruhe Institute of Technology\\
Wolfgang-Gaede-Str.~1, 76131 Karlsruhe, Germany}

\maketitle\abstracts{
The Brout-Englert-Higgs mechanism describes the generation
of masses of fundamental particles in the Standard Model~(SM).
It predicts the existence of one scalar particle
with precisely predicted couplings to fermions and gauge bosons.
Deviations from these predictions, such as the observation of
additional scalar particles, would indicate non-minimal Higgs
sectors and beyond-the-SM physics. We motivate extended scalar
sectors based on theoretical and observational grounds and
discuss their possible manifestations at colliders or
gravitational-wave experiments. We explore smoking-gun
signatures of electroweak baryogenesis at the LHC or a
future $e^+e^-$ collider, and we examine their interplay with the
observation of a primordial gravitational-wave
background at LISA. We also consider the possibility of
an additional Higgs boson at about 95~GeV motivated by
excesses in diphoton and ditau final states
observed by CMS. Finally, we discuss the impact of the
CDF measurement of the $W$-boson mass on the parameter
space of scalar extensions of the SM.}

\section{Introduction}
\label{sec:intro}

The Standard Model~(SM) of particle physics has
shown remarkable accuracy in predicting observations
at high-energy collider experiments, particularly at
the Large Hadron Collider (LHC), over a broad range
of energy scales. The Brout-Englert-Higgs
mechanism~\cite{Englert:1964et,Higgs:1964pj} of
the electroweak (EW) theory describes the generation of
particle masses via spontaneous breaking of the EW symmetry.
This symmetry breaking results from the non-zero vacuum expectation
value (vev) of the Higgs field. The SM incorporates a
minimal prescription of EW symmetry-breaking featuring
a single scalar SU(2) doublet field. This yields  three
key predictions which are currently tested at the LHC:
(i) the existence of a single scalar particle which, within the current
experimental precision, can be identified
with the detected Higgs boson at about 125~GeV,
denoted $h_{125}$ in the following,
(ii) its coupling strengths to massive particles are
determined by their masses and they increase with increasing mass, and
(iii) no sources of CP violation in the Higgs potential.

Although the SM provides a parametrisation of EW
symmetry breaking, it does not describe the underlying
physics. 
Moreover, the SM's Higgs sector faces the problem that
the Higgs-boson mass is not protected by a symmetry
from larger energy scales and is instead quadratically
dependent on any large physical scale present in nature.
Thus, a light Higgs boson is considered highly unnatural
and not well-understood theoretically.

These shortcomings of the SM can be addressed in beyond
the~SM~(BSM) theories.
A wide class of BSM theories feature extended scalar sectors
in which additional scalar particles are present at
the EW scale. One of the key tasks of the current and
future LHC programme is to search for these
additional scalar particles to shed light on the
physics underlying EW symmetry breaking.
Extended scalar sectors also allow for the possibility
that the detected Higgs boson
is incorporated in such a way that it only
approximately resembles a SM Higgs boson.
As a result, the presence of extended Higgs sectors
is also probed by the LHC mass and coupling measurements
of $h_{125}$. Given its precisely predicted nature in the SM,
$h_{125}$ is an ideal probe of new physics, since
any modifications of its signal rates would
clearly indicate physics beyond the SM. 

The SM has also phenomenological shortcomings that
motivate the consideration of extended
Higgs sectors. One of the most pressing open issues
is the matter-antimatter asymmetry of the universe,
which cannot be explained in the SM~\cite{Kajantie:1996mn}.
A possible explanation for the observed matter-antimatter
asymmetry is EW baryogenesis~\cite{Kuzmin:1985mm},
where a baryon asymmetry is generated during a
first-order EW phase transition through sphaleron processes.
However, to render the EW phase transition to be
of first order, the SM has to be extended by new
degrees of freedom at energy scales close to the EW scale.
The addition of BSM scalar fields that themselves could
obtain vevs is the most convincing possibility.
If the transition is sufficiently strong, the
sphalerons are highly suppressed in the EW symmetry breaking
phase after the transition such that the previously generated
asymmetry cannot be washed out anymore until present times.
Given the sizable modifications of the Higgs sector compared
to the SM, models that can realize EW baryogenesis can be probed
by the LHC, making EW baryogenesis a particularly interesting
solution to the unknown origin of the matter-antimatter asymmetry.

A new window to the epoch of the EW
phase transition will open once the space-based
gravitation wave detector LISA will start taking data.
LISA is designed to detect
gravitational waves in a frequency band that
matches those of the stochastic gravitational
wave background as it would be generated during
a first-order EW phase
transition~\cite{LISACosmologyWorkingGroup:2022jok}.
Thus, the exploration of the EW scale during the next
decades will be marked by the complementarity between
the future high-luminosity Runs of the LHC and the
LISA experiment.

Upon extending the SM Higgs sector, important
experimental restrictions have to be considered.
Most importantly, the EW
$\rho$-parameter~\cite{Ross:1975fq} defined by
$\rho = M_W^2 / (\cos^2\theta_w M_Z^2)$
has beend measured experimentally to have
a value of $\rho_{\rm exp} = 1.00038 \pm 0.00020$~\cite{ParticleDataGroup:2022pth}.
As a consequence of the custodial symmetry,
the SM predicts $\rho = 1$ at the clasical level,
in remarkable agreement with $\rho_{\rm exp}$.
If one adds additional EW multiplets to the
Higgs sector of the SM, the prediction for the $\rho$-parameter
is in general different from one.
It is therefore convincing to study models in
which $\rho \approx 1$ holds without fine tuning of
the model parameters, such that the experimentally
measured value is not just
a strange coincidence.
It turns out that the addition of only gauge singlet
scalar fields and SU(2) doublet fields maintains
the SM prediction
$\rho = 1$ at the classical level. We will therefore focus in
the following on the two Higgs doublet model~(2HDM)
and the singlet-extended 2HDM~(S2HDM).

\section{The (singlet-extended) two Higgs doublet model}
\label{sec:models}

The two
Higgs doublet model (2HDM) 
contains two scalar SU(2) doublet fields
$\Phi_1$ and $\Phi_2$~\cite{Lee:1973iz}.
Neglecting sources of CP violation,
and assuming the presence of a softly broken $Z_2$ symmetry under which
one of the doublet fields changes its sign, the scalar potential
can be written as
\begin{eqnarray}
V_{\rm 2HDM} &= \;
m_{11}^2 |\Phi_1|^2 + m_{22}^2 |\Phi_2|^2 - m_{12}^2 (\Phi_1^\dagger
\Phi_2 + \mathrm{h.c.}) + \frac{\lambda_1}{2} (\Phi_1^\dagger \Phi_1)^2 +
\frac{\lambda_2}{2} (\Phi_2^\dagger \Phi_2)^2 \nonumber \\
&+ \lambda_3
(\Phi_1^\dagger \Phi_1) (\Phi_2^\dagger \Phi_2) + \lambda_4
(\Phi_1^\dagger \Phi_2) (\Phi_2^\dagger \Phi_1) + \frac{\lambda_5}{2}
[(\Phi_1^\dagger \Phi_2)^2 + \mathrm{h.c.}] \ . 
\label{eq:scalarpot}
\end{eqnarray}
The $Z_2$ symmetry can be extended to the Yukawa sector, where
as a result of the opposite $Z_2$ parities the scalar doublets
cannot be coupled both to the same fermion. This leads
to the absence of flavour-changing neutral currents at
classical level in agreement with experimental observations.
Depending on
the different possibilities for assigning $Z_2$ parities to
the different kind of fermions, one finds
the four different Yukawa types~I,
II, III~(lepton-specific) and IV~(flipped).

The two Higgs doublets have a total of eight degrees
of freedom, from which three belong to the Goldstone
bosons after EW symmetry breaking and are thus
absorbed into the longitudinal degrees of freedom
of the massive gauge bosons $W^\pm$ and $Z$.
The remaining degrees of freedom give rise to five
physical scalar states: two scalar states $h$ and
$H$, where in the following $h$ is assumed to be
the lighter state, and which plays the role of the detected
Higgs boson at 125~GeV, a pseudoscalar state $A$,
and a pair of charged Higgs bosons $H^\pm$.
The mixing of $h$ and $H$
is commonly expressed in terms of the
mixing angle~$\alpha$.
In the \textit{alignment limit}
$\cos(\beta - \alpha) = 0$, where the parameter $\beta$
is defined by $\tan\beta = v_2 / v_1$, the couplings of
the state $h$ are equal to the couplings of
a SM Higgs boson.

Extending upon the 2HDM, the
singlet extended~2HDM~(S2HDM)
adds an additional complex scalar
field $\Phi_S$ acting as a singlet
under the gauge symmetries.
We will consider a variation of the model in which
$\Phi_S$ transforms under an additional global
U(1) symmetry, whereas all other fields are invariant
under this symmetry~\cite{Biekotter:2021ovi}.
We furthermore assume
the spontaneous breaking of the U(1) symmetry by means of a non-zero
vev $v_S$ of the singlet field. To avoid
the presence of a massless mode as a Goldstone boson
associated with this symmetry breaking, we assume
the presence of a bilinear U(1)-breaking term in
the potential. Accordingly, the scalar potential
can be written as
\begin{eqnarray}
V_{\rm S2HDM} = V_{\rm 2HDM} +
\frac{1}{2} m_S^2 |\Phi_S|^2 + \frac{\lambda_6}{2} |\Phi_S|^4 +
\frac{\lambda_7}{2} |\Phi_1|^2 |\Phi_S|^2 +
\frac{\lambda_8}{2} |\Phi_2|^2 |\Phi_S|^2
- \frac{\mu_\chi^2}{4} ( \Phi_S^2 + \mathrm{h.c.} ) \ ,
\label{eq:scalarpot2}
\end{eqnarray}
where the terms proportional to $\mu_\chi^2$ softly break
the global U(1) symmetry. As before, the invariance under
the discrete $Z_2$ acting on $\Phi_1$ and $\Phi_2$ was imposed
to suppress tree-level flavour-changing neutral currents.

The S2HDM has two additional physical scalar states
in comparison to the 2HDM. The real component of $\Phi_S$
mixes with the neutral real compoments of $\Phi_1$ and
$\Phi_2$, forming a total of three CP-even Higgs bosons $h_{1,2,3}$.
The imaginary component of $\Phi_S$ gives rise to a massive
stable scalar state $\chi$ that can act as a dark-matter
candidate in the form of pseudo-Nambu-Goldstone dark matter.
This form of Higgs-portal dark matter gained attention in
recent years since the cross sections for the scattering
of $\chi$ on nuclei are vanishing at classical level due
to a cancellation mechanism resulting from the global
U(1) symmetry. As a result, the S2HDM passes
constraints from dark-matter direct-detection
experiments without a tuning of
parameters~\cite{Biekotter:2022bxp}.
In total analogy to the 2HDM, the scalar spectrum contains
the pseudoscalar state $A$ and the charged Higgs bosons $H^\pm$.

\section{Experimental opportunities}
\label{sec:oppo}

As already mentioned above, so far no clear hints
for extended scalar sectors (or any other kind
of BSM physics) have been observed at the LHC.
In the following, I will discuss several possibilities
for which the presence of an extended Higgs sector
could still reveal itself in the near future at the LHC.
I will also discuss the possible interplay with
other experiments, such as the LISA experiment, either
in order to further constrain the parameter
space of BSM theories, or in order to discriminate
between different BSM theories if evidence for
BSM physics will be found in the near future.

\subsection{Smoking-gun signals of electroweak
baryogenesis}

Under the presence of a second Higgs doublet,
it is possible to render the EW phase transition
to be of strongly first-order, which is a vital
ingredient for the realization of EW
baryogenesis.
In the type~II 2HDM, the most promising
scenario for the realization of a first-order
phase transition is characterized by the mass
hierarchy $m_H \ll m_A \approx m_{H^\pm}$. 
Due to the mass splitting between~$H$ and~$A$, the process
$A \to Z H$ has been coined as a smoking-gun
signature of~EW baryogenesis in the 2HDM~\cite{Dorsch:2014qja}.
Both ATLAS and CMS have searched for this
signature assuming that $H$ decays into
bottom-quark or tau-lepton pairs.
In Ref.~\cite{Biekotter:2022kgf} we have shown
that as a consequence of the existing constraints
on the type~II 2HDM, $H$ tends to be
so heavy that it is predicted to dominantly decay
into top-quark pairs. Notably, results for the search for
the smoking-gun signature in $Z t \bar t$ final states
has not been published yet, but efforts to analyze
such final states are ongoing by both ATLAS and CMS.

\begin{figure}[t]
\centering
\includegraphics[width=0.4\textwidth]{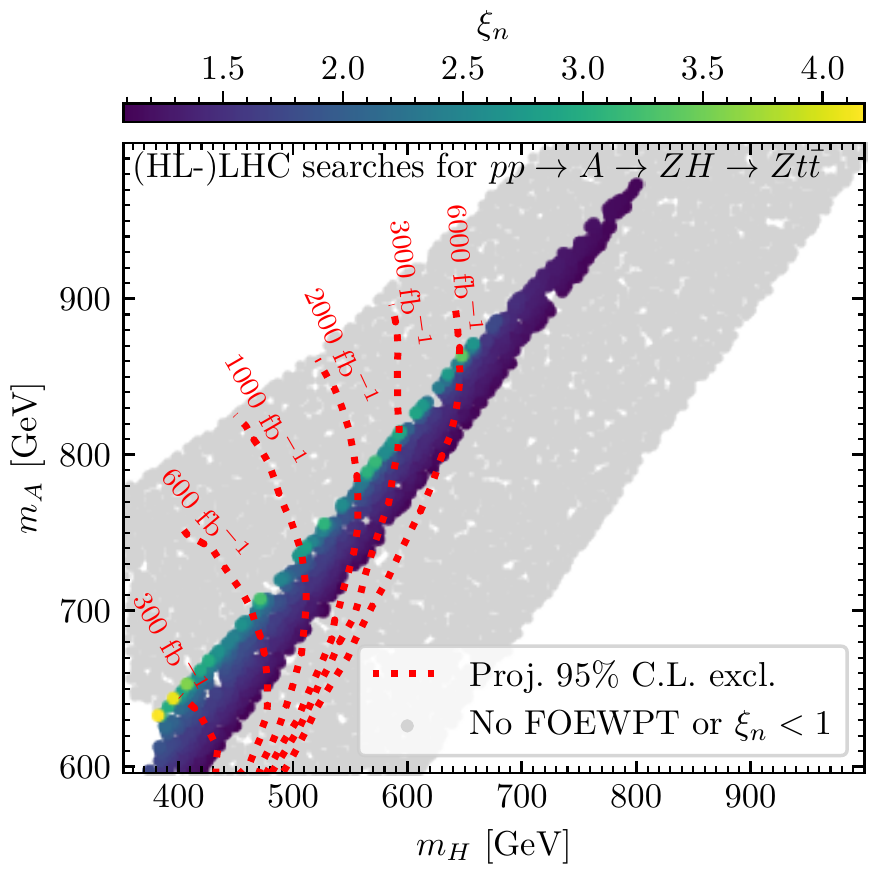}
\caption{
The $(m_H,m_A)$ plane in the type~II 2HDM
for $\tan\beta = 3$, $\cos(\alpha - \beta) = 1$,
 and $m_{12}^2 = m_H^2 \sin\beta \cos\beta$,
with the color coding indicating the value of
$\xi_n$ if $\xi_n > 1$.
The remaining points are shown in grey.
The red dashed lines indicate the
projected 95\% confidence-level exclusion
regions resulting
from the (HL-)LHC searches for the process
$pp \to A \to ZH$ with $H$ decaying into a
pair of top quarks.
Taken from Ref.~\protect\cite{Biekotter:2022kgf}.}
\label{sg1}
\end{figure}

In Fig.~\ref{sg1} we show the results of a scan
in the alignment limit of the type~II 2HDM
for $\tan\beta = 3$~\cite{Biekotter:2022kgf}.
Here the parameter points featuring a strong
first-order phase transition are indicated
with the colors. The color coding indicates
the value of $\xi_n = v_n / T_n$, where $v_n$
and $T_n$ are the vev $v$ and the temperature
$T$ at the transition, respectively, and where a strong
transition is defined
by $\xi_n > 1$, as such transition prevents any
washout of the baryon asymmetry.
The remaining points
are shown in grey.
We also show the expected 95\%
confidence-level exclusion sensitivity for different
values of the integrated luminosity $\mathcal{L}$.
These were obtained from a naive rescaling of
expected CMS limits assuming
$\mathcal{L} = 41~\mathrm{fb}^{-1}$
(see Ref.~\cite{Biekotter:2022kgf} for details).
Comparing the red dashed lines to the region
in which a first-order phase transition can be
realized, one can see that the LHC will have
a significant discovery potential, in particular
during its high-luminosity phase.
On the other hand, if no deviations from the
SM background expectation will be found,
large parts of the 2HDM parameter regions suitable
for EW baryogenesis would be excluded.
Here it should be taken into account that
the gluon-fusion production cross section
of~$A$ scales with $1 / (\tan\beta)^2$, such that
for smaller values of $\tan\beta$, as preferred
for a realization of EW baryogeneis,
the experimental prospects
are even better.\footnote{We 
note that the measurement of the cross
section for this signature might reveal important
information about the underlying model that
is realized in nature. For instance,
possible ways to distinguish between
a 2HDM and an S2HDM realization by
means of the $A \to ZH$ signature
have been discussed in Ref.~\cite{Biekotter:2021ysx}.}

\begin{figure}[t]
\centering
\includegraphics[width=0.7\textwidth]{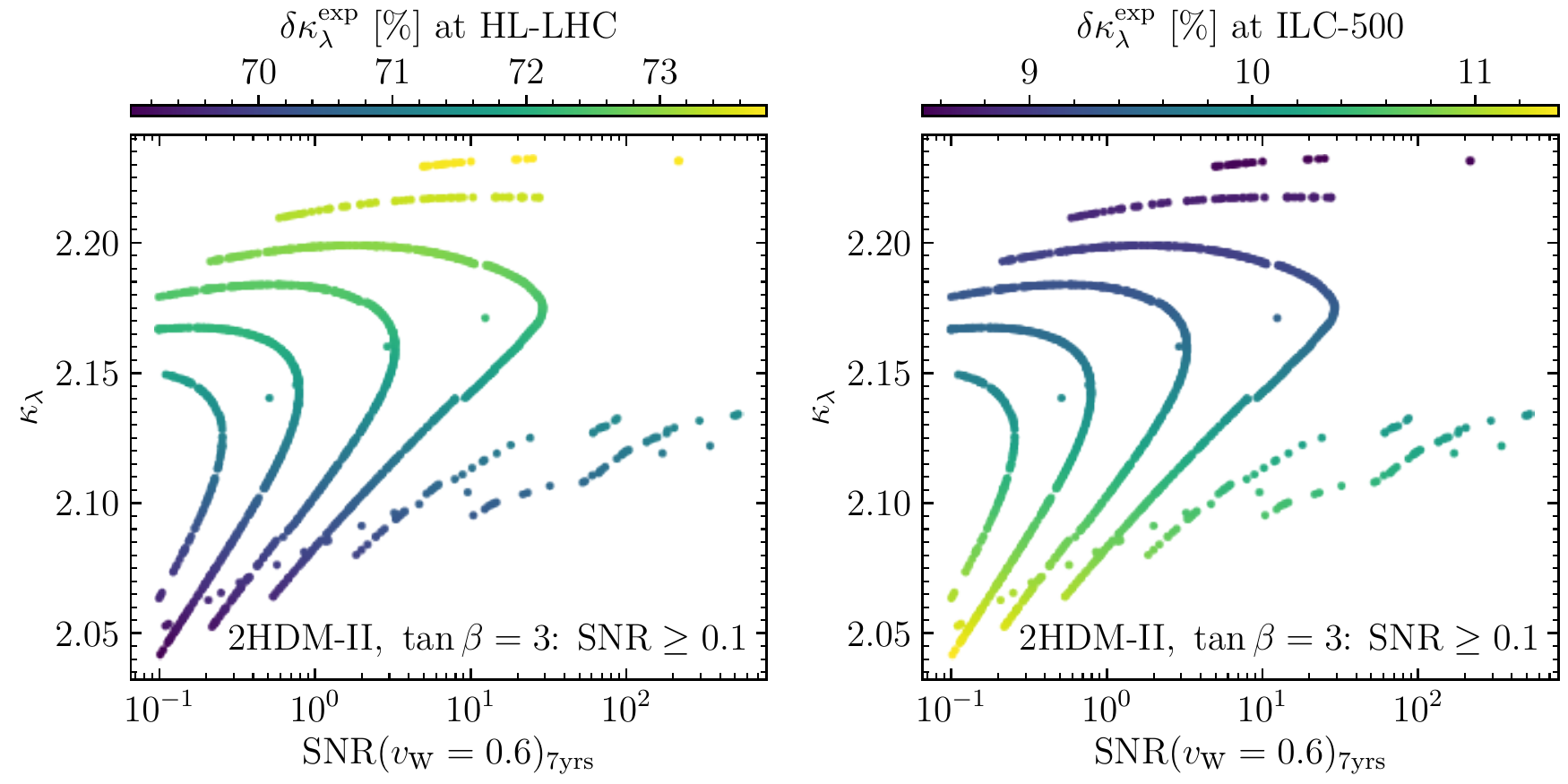}
\caption{
2HDM type~II parameter points with
$\mathrm{SNR} \geq 0.1$ in the
(SNR, $\kappa_\lambda$) plane.
The color coding of the points indicates the
projected experimental precision of the
measurement of $\kappa_\lambda$ at
the HL-LHC (left) and the ILC-500 (right),
see text for details.
Taken from Ref.~\protect\cite{Biekotter:2022kgf}.}
\label{sg2}
\end{figure}

A first-order EW phase transition in the 2HDM
is not only associated with  BSM Higgs bosons
not much heavier than the EW scale, but it
is also linked to sizable modifications of
the trilinear self-coupling of~$h = h_{125}$.
This modifications are usually expressed
in terms of the parameter $\kappa_\lambda =
\lambda_{hhh} / \lambda_{hhh}^{\rm SM}$,
where $\lambda_{hhh}$ is the 2HDM coupling prediction,
and $\lambda_{hhh}^{\rm SM}$ is the tree-level
Higgs boson self-coupling in the SM.
In Fig.~\ref{sg2}, in which we depict the predictions
for $\kappa_\lambda$ against the signal-to-noise
ratio~(SNR) at LISA for the detection of the primordial
gravitational-wave background as it would have
been generated during the phase
transition~\cite{Biekotter:2022kgf}.
We only show here the predictions for the subset
of parameter points shown in \reffi{sg1} for which
the SNR is of the order of one or larger, i.e.~for
parameter points that feature a gravitational-wave
signal that potentially is detectable at LISA.\footnote{See
Ref.~\cite{Biekotter:2022kgf} for details on the prediction of
the gravitational-wave signals.}
One can see that these points predict $\kappa_\lambda
\approx 2$.
The LHC is expected to be able to disfavour
such values of $\kappa_\lambda$ at the level
of $2\sigma$ if no deviations from the
SM expectations for the non-resonant
pair-production of~$h_{125}$
will be found during the high-luminosity
phase~\cite{ATL-PHYS-PUB-2022-018}.
Here one should note that the destructive interference
between the $s$-channel and the box diagram
leads to a smaller LHC cross section for the pair-production
of~$h_{125}$ for values of $\kappa_\lambda \approx 2$
compared to the SM prediction. As a result, assuming
that $\kappa_\lambda \approx 2$ is realized in nature,
the relative precision with which $\kappa_\lambda$ could be
determined at the LHC (indicated with the colors of
the points in the left plot of Fig.~\ref{sg2})
is only about $70\%$. As a comparison, the expected
precision on $\kappa_\lambda \approx 1$, i.e.~assuming
SM cross sections for $h_{125}$-pair production,
is $60\%$. In the right plot of Fig.~\ref{sg2}, the color
coding indicates the relative precision of a
measurement of $\kappa_\lambda$ at the International
Linear Collider (ILC) operating at $\sqrt{s} = 500\gev$.
One can see that an $e^+e^-$ collider with sufficient energy
to pair-produce $h_{125}$ could improve substantially
the experimental precision with which $\kappa_\lambda$-values
of about two could be determined.

The results discussed in this section demonstrate
that the Higgs measurements at the LHC in the upcoming
years will shape in a profound way the expectations
for the possibility
of measuring a gravitation-wave background produced
during a first-order EW phase transition at LISA.
In the hypothetical scenario in which no additional Higgs bosons
will be found, and in which the coupling measurements of the
detected Higgs boson will show no indications of
BSM physics, the LHC will limit the possibility of
detecting a primordial gravitational-wave background as a remnant
of the EW phase transition to corners of the parameter
space of many models.

\subsection{A Higgs boson at 95~GeV?}
\label{sec:95}

Many phenomenological analyses of models
with extended scalar sectors focus on scenarios
in which $h_{125}$ is the lightest Higgs boson.
However, additional Higgs bosons lighter than
125~GeV are still experimentally viable if their
couplings to gauge bosons are suppressed compared
to the ones of a SM Higgs boson, for instance
due to a large singlet admixture.

In view of this, it is interesting that several
local excesses have been observed at a mass of about
95~GeV: (i) LEP observed an excess in $e^+e^- \to Z H$
with $H$ decaying into bottom-quark pairs
($2.3\sigma$ local significance)~\cite{Barate:2003sz}.
(ii) CMS observed an excess in $p p \to H \to \tau^+ \tau^-$
using the full Run~2 dataset ($3.1\sigma$ local
significance)~\cite{CMS:2022goy}. (iii) CMS observed an excess in
$p p \to H \to \gamma \gamma$ in the 8~TeV dataset
($2\sigma$ local significance)~\cite{CMS:2015ocq} and
in the 13~TeV dataset ($2.9\sigma$ local
significance)~\cite{Sirunyan:2018aui,CMSnew}.
The latter diphoton excess at 13~TeV was reported
previously based only on the first-year Run~2
dataset~\cite{Sirunyan:2018aui}.
During the Moriond conference, the updated result
including the full Run~2 data was presented~\cite{SusanTalkHD}, showing
a practically unchanged significance of the excess,
but the excess corresponds to a smaller signal rate
as compared to the previous result.
It should be noted that,
in addition to the inclusion of more data,
the updated result~\cite{CMSnew} also has a refined experimental
analysis for the
background rejection of misidentified
$Z \to e^+ e^-$ Drell-Yan events, and
further event classes requiring
additional jets have been utilized.
Taking this into account, from our point of view,
the persistence of the diphoton excess
strengthens the motivation to investigate a
BSM interpretation of the excesses.

\begin{figure}[t]
\centering
\includegraphics[width=0.34\textwidth]{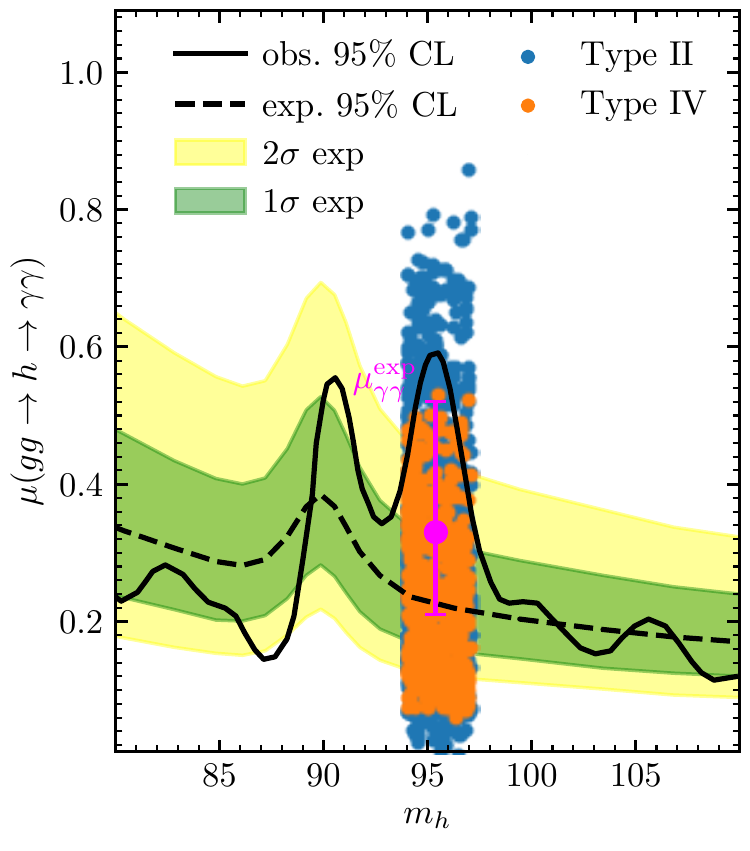}
\caption{
S2HDM parameter points
passing the applied constraints
in the $(m_{h_{95}},\mu_{\gamma\gamma})$
for the type~II (blue) and the type~IV (orange).
The expected and observed
cross section limits obtained
by CMS are indicated by
the black dashed and solid lines, respectively,
and the $1\sigma$ and $2\sigma$ uncertainty intervals
are indicated by
the green and yellow
bands, respectively. The value of
$\mu_{\gamma\gamma}^{\rm exp}$ and its uncertainty
is shown with the magenta error bar at the mass
value at which the excess is most pronounced.
Taken from Ref.~\protect\cite{Biekotter:2023jld}.}
\label{951}
\end{figure}

In Ref.~\cite{Biekotter:2023jld} we have investigated
whether the S2HDM can accommodate the excesses by means
of a singlet-like scalar at 95~GeV whose
couplings to the fermions and gauge bosons result from
the mixing with $h_{125}$. As shown in Fig.~\ref{951},
both type~II and type~IV (flipped) can give rise to
a state that describes the CMS diphoton excess, while
being in agreement with all relevant theoretical
and experimental constraints.
In our previous studies~\cite{Biekotter:2019kde,Biekotter:2021ovi,Biekotter:2022jyr,Biekotter:2022abc},
in which a signal rate
for the diphoton excess of $\mu_{\gamma\gamma}^{\rm exp} = 0.6 \pm 0.2$
based on the earlier CMS analysis including only the first-year
Run~2 dataset was considered~\cite{Sirunyan:2018aui}, we found a preference for
type~II, in which larger diphoton signal rates can be achieved.
The updated value~\cite{SusanTalkHD} of $\mu_{\gamma\gamma}^{\rm exp} =
0.33_{-0.12}^{+0.19}$
is such that both type~II and type~IV can describe the diphoton
excess equally well. General implications of the reduction of
$\mu_{\gamma\gamma}^{\rm exp}$ for an interpretations in
different classes of BSM theories are discussed in
Ref.~\cite{Biekotter:2023jld}.

We have also analysed
whether the other
excesses at about 95~GeV can be described in addition
to the diphoton excess. We have shown that both type~II
and type~IV can additionally accommodate the LEP excess,
however only in type~IV also a sizable signal for the
CMS ditau excess can be achieved.
However, even in type~IV the ditau excess
can only be described at the level of $1\sigma$
due to constraints from a related CMS search for
$pp \to tt H$ with $H \to \tau^+ \tau^-$~\cite{CMS-PAS-EXO-21-018},
in which no excess has been observed at and around 95~GeV.

\subsection{The CDF W-boson mass measurement and
isospin splitting}

The SM does not predict the masses of the
fundamental particles. However, one can
find relations between
the masses and other measured quantities.
The mass of the $W$ boson $M_W$ can
be expressed 
from muon decay
as a function of the Fermi constant
$G_\mu$, the fine structure constant $\alpha$,
and the masses of the $Z$ boson
via the relation
\begin{equation}
M_W^2 = M_Z^2 \left(
\frac{1}{2} +
\sqrt{
  \frac{1}{4} -
  \frac{\alpha \pi}{\sqrt{2} G_\mu M_Z^2}(1 +
    \Delta r)}
\right) \ . 
\end{equation}
Here, the parameter $\Delta r$ incorporates the
quantum corrections, which in the SM have been
evaluated up to four-loop order.
It is important to note that the leading order
prediction for $M_W$ in the SM
is in large disagreement with the experimentally
measured value, and only the inclusion of the quantum
corrections shift the prediction to larger values
in agreement with the 2022 PDG average value.
One of the most important corrections contained in
$\Delta r$ is given by the EW $\Delta \rho$-parameter
which quantifies the breaking of the custodial symmetry
(see the discussion in Ref.~\ref{sec:intro}).
In the SM, the most important corrections contained
in $\Delta \rho$ result from the isospin splitting
in the fermion sector, where due to the large mass splitting
between top quark and bottom quark one finds the
approximate proportionality $\Delta \rho \sim (m_t^2 - m_b^2)$.

In the year 2022 the CDF collaboration has published
a new measurement of the $W$-boson mass utilizing
an integrated luminosity of $8.8~\mathrm{fb}^{-1}$ collected
at the Tevatron $p \bar p$ collider. The result
was $M_W^{\rm CDF} = 80.4335 \pm 0.0094$~GeV, which deviates
from the SM prediction for $M_W$ by about $7 \sigma$.
However, one should have in mind that the CDF measurement
is not only in disagreement with the~SM, but also with all
previous measurements of $M_W$ at LEP, the Tevatron and
the~LHC. At this moment in time, one cannot disregard
neither the CDF measurement (being the most precise one)
nor the previous measurements of $M_W$ (being consistent
within each other and with the SM).\footnote{The ATLAS
collaboration reported an update on their $M_W$ determination
during this year's Moriond EW
conference~\cite{ATLAS-CONF-2023-004}, where
an improved statistical interpretation was applied,
showing good agreement with the SM.}

\begin{figure}[t]
\centering
\includegraphics[width=0.4\textwidth]{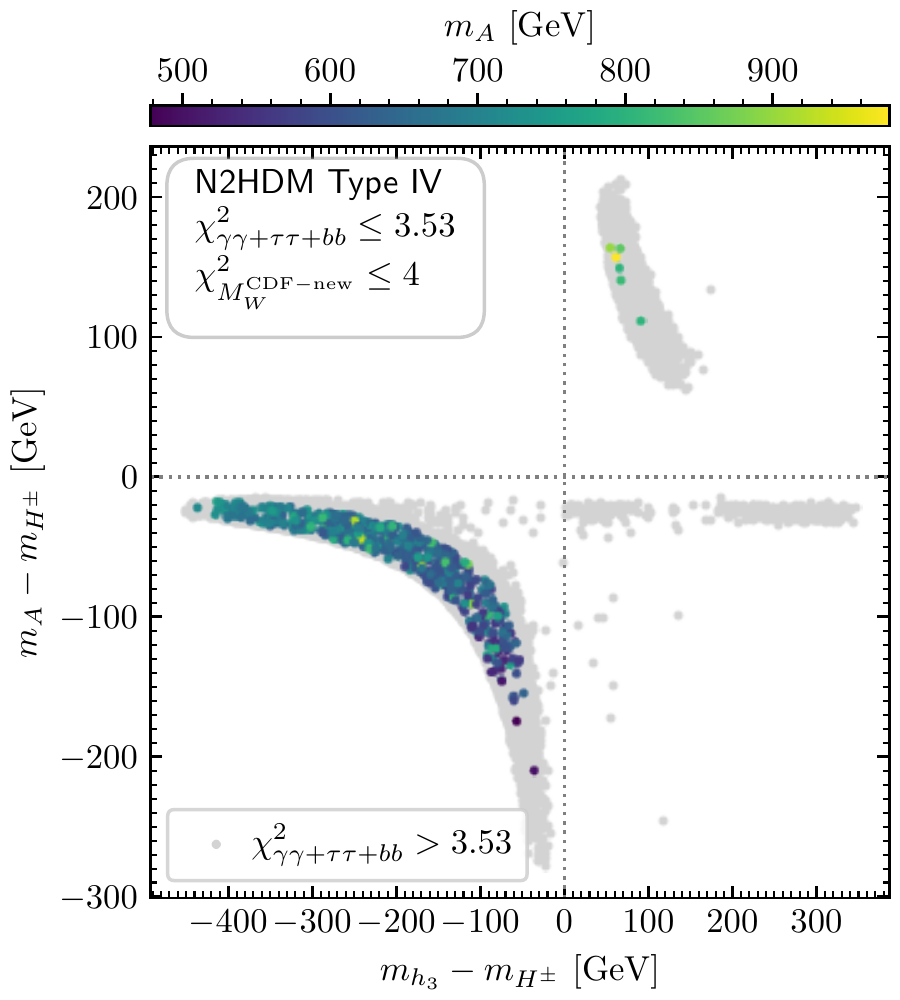}~
\includegraphics[width=0.4\textwidth]{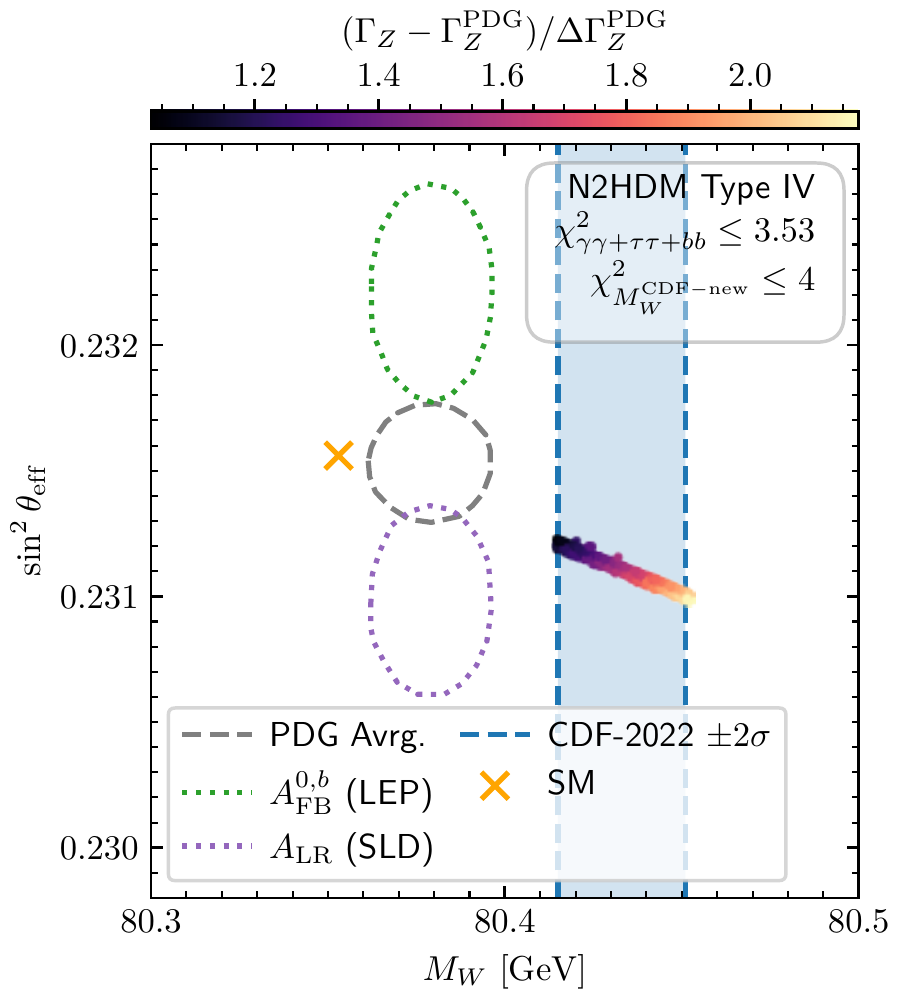}
\caption{Left:
S2HDM parameter points that predict a value of $M_W$
within the $2\sigma$ uncertainty band of
the CDF measurement in the plane of the mass
differences $m_{h_3} - m_{H^\pm}$
and $m_{A} - m_{H^\pm}$.
The color coding indicates the value of
$m_A$ for the parameter points
that describe the excesses at
$95\gev$ at the level of $1 \ \sigma$
or better.
The remaining parameter points, are shown in gray.
Right:
The predictions for $M_W$ and $\sin^2\theta_{\rm eff}$ in the S2HDM.
The color coding of the points indicates the
difference between the prediction for
$\Gamma_Z$ and the PDG average value
$\Gamma_Z^{\rm PDG}$ divided by the
experimental uncertainty
$\Delta \Gamma_Z^{\rm PDG}$.
The light blue region corresponds
to the new CDF measurement within $\pm 2 \ \sigma$.
The violet and the green dotted
ellipses indicate
the $68\%$ confidence level limits from
the two individually most precise 
measurements of $\sin^2\theta_{\rm eff}$ via
$A_{\rm FB}^{0,b}$ at LEP and
$A_{\rm LR}$ at SLD, respectively,
whereas the gray dashed ellipse
indicates the PDG average
The orange cross indicates the SM prediction.
Both plots taken from Ref.~\protect\cite{Biekotter:2022abc}.}
\label{mw1}
\end{figure}

Nevertheless, given the large discrepancy with the SM, one can ask
the question whether there are models with extended scalar
sectors which could accommodate a sizable upwards shift
to $M_W$, even as large as the CDF result, without being
in tension with other observables.
Scalar extensions with a second SU(2) Higgs doublet
are promising candidates, since the additional BSM
scalars can give rise to additional sources of
isospin splitting. In the 2HDM, one finds at the
one-loop level that $\Delta \rho \sim (m_A^2 - m_{H^\pm}^2)
(m_H^2 - m_{H^\pm}^2)$, i.e.~an upwards shift to
the prediction for $M_W$ requires the pseudoscalar
$A$ and the heavy scalar $H$ to be both lighter
or both heavier than the charged scalars $H^\pm$.
In Ref.~\cite{Biekotter:2022abc} we exploited this
feature in order to simultaneously accommodate the
CDF $M_W$ measurement and the collider excesses at~95GeV
in the S2HDM, see left plot of Fig.~\ref{mw1}.
In the right plot of Fig.~\ref{mw1} one can see that
in this case also the prediction for the effective
weak mixing angle $\sin^2\theta_{\rm eff}$ is
affected, making it more
compatible with the SLD measurement of $A_{\rm LR}$,
but less so with the LEP measurement of $A_{\rm FB}^{0,b}$.
No significant tension for the width
of the $Z$ boson $\Gamma_Z$ is found.

\section{Conclusions}
\label{sec:conclu}

After the discovery of a SM-like Higgs boson
at 125~GeV, a prime goal of the current and
future LHC science programme is to shed more
light on the physics underlying EW symmetry
breaking. No clear evidence for physics beyond
the~SM has been found yet, but models with
extended Higgs sectors are well motivated candidates.
Here we discussed several measurements by which
the presence of an extended Higgs sector might
reveal itself in the near future.

\section*{Acknowledgments}
I thank my collaborators with whom the
results reviewed here have been obtained:
P.~Gabriel, S.~Heinemeyer, M.O.~Olea Romacho,
J.M.~No, R.~Santos and G.~Weiglein.
I furthermore
thank M.O.~Olea Romacho for valuable comments
during the preparation of my talk.
I am grateful to the organizers of the Moriond conference
for the kind invitation, the welcoming atmosphere
during the conference, and for giving me the opportunity
to present my work.
This work is supported by the German
Bundesministerium f\"ur Bildung und Forschung
(BMBF, Federal Ministry of Education and Research)
-- project 05H21VKCCA.

\section*{References}

\bibliography{refs}

\end{document}